# An Automatic Architecture Reconstruction and Refactoring Framework

**Frederik Schmidt, Stephen G. MacDonell, Andrew M. Connor**

SERL, Auckland University of Technology
Private Bag 92006, Auckland 1142
New Zealand
+64 9 921- 8953

{fschmidt, stephen.macdonell, andrew.connor}@aut.ac.nz

**Abstract**   A variety of sources have noted that a substantial proportion of non trivial software systems fail due to unhindered architectural erosion. This design deterioration leads to low maintainability, poor testability and reduced development speed. The erosion of software systems is often caused by inadequate understanding, documentation and maintenance of the desired implementation architecture. If the desired architecture is lost or the deterioration is advanced, the reconstruction of the desired architecture and the realignment of this desired architecture with the physical architecture both require substantial manual analysis and implementation effort. This paper describes the initial development of a framework for automatic software architecture reconstruction and source code migration. This framework offers the potential to reconstruct the conceptual architecture of software systems and to automatically migrate the physical architecture of a software system toward a conceptual architecture model. The approach is implemented within a proof of concept prototype which is able to analyze java system and reconstruct a conceptual architecture for these systems as well as to refactor the system towards a conceptual architecture.

## 1 Introduction

Software systems evolution is triggered by changes of business and technical requirements, market demands and conditions [1]. This evolution leads often to an erosion of the architecture and design of the system [2]. Reasons for this are several – insufficient time for design improvements [3], no immediate architecture maintenance and management, fluctuation of employees, and different levels of knowledge and understanding of the conceptual architecture. Additionally, the automatic inclusion of imports within current IDE's obfuscates the creation of unwanted dependencies which conflict with the conceptual architecture. This erosion of design leads to fragile, immobile, viscous, opaque and rigid software systems [4]. Architecture reconstruction tools can help to create views of the physical architecture of the system [5], however these views do not fully support the ability to reconstruct a con-ceptual architecture view of the system which can be utilized as a blueprint for further development. Architecture management tools [6] such as *Lattix, Sotograph* and *SonarJ* help to monitor the design drift. To apply these, however, a clear conceptual architecture has to be defined. These tools offer only limited support to realign the conceptual and physical architectures [6]. Manual recreation of the conceptual architecture is hindered as the design erosion obfuscates the original intent of the system design. Additionally, existing architecture violations introduced during the system erosion have to be resolved to realign the physical architecture with the conceptual architecture. To avoid, or at least simplify, this complex and time consuming manual process we introduce an automatic framework for reconstruction and refactoring. This framework features the reconstruction of a conceptual architecture model based on acknowledged design principles and the resolution of architecture violations by applying software migration techniques. This facilitates the migration of the physical architecture model towards the conceptual architecture model. Before the framework is described, related work on software architecture reconstruction and migration is presented.

## 2 Related Work

The objective of this work relates to a variety of software engineering disciplines such as software architecture modeling, software architecture reconstruction, software architecture transformation and automatic refactoring of legacy systems. Strategies of software clustering and search based software engineering are applied to realise the objective of our study. Additionally, software architecture metrics are applied to evaluate the value of the generated solutions. To highlight the relevance and significance of our study relevant and contributing work of the named areas are illustrated in this section.

### *2.1 Software Architecture*

The architecture of a software system visualizes a software system from an abstract view. An architectural view of a system raises the level of abstraction, hiding details of implementation, algorithms and data representations [7]. These architectural views can focus on different aspects of the system for example a service, implementation, data or process perspective [5]. Associated fine grained entities are classified into more abstract modules. Having a current representation of the system architecture is crucial in order to maintain, understand and evaluate a large software application [8].

Murphy & Notkin [9] depict the reflexion model as a framework to prevent software architecture deterioration of the implementation perspective. The reflexion model features a conceptual architecture which defines a desired model of the system and a mapping of the physical implementation units into the subsystems of the conceptual architecture model. The conceptual architecture model facilitates a development blueprint for the development stakeholders. An ideal conceptual architecture models the domain and technical environment of the software system and delivers a framework to maintain desired quality aspects.

The approach adopted by Murphy & Notkin [9] demands regular compliance checking of the physical and conceptual architectures and the immediate removal of architecture viola-tions by applying refactoring techniques [10]. Therefore the documentation of a conceptual architecture and compliance checking of the conceptual and physical architecture is an important aid to maintain and understand software systems [11]. Within many software projects the compliance checking of physical architecture and conceptual architecture as well as the inclusion of realignment of the physical and conceptual architecture is not consequently included into the development process [12]. Additionally, software systems evolve due to requirement and framework changes during development. This may require altering the conceptual architecture. Consequently, the conceptual and physical architecture drifts apart without a rigorous compliance checking and refactoring of the physical architecture. The manual reconstruction of the architecture as well as the manual realignment is complex and time consuming.

## 2.1 Software Architecture Reconstruction

One of the challenges associated with architecture reconstruction is that often the available documentation is incomplete, outdated or missing completely. The only documentation of the software system is the source code of the system itself.

Reverse engineering activities help to obtain abstractions and views from a target system to help the development stakeholders to maintain, evolve and eventually re-engineer the architecture. The main objective of software architecture reconstruction is to abstract from the analyzed system details in order to obtain general views and diagrams with different metrics associated with them. Even if the source code might be eroded it is often the only and current documentation of the software system. Therefore the extraction of architecturally significant information and its analysis are the key goals which have to be determined to apply software architecture reconstruction successfully.

A variety of approaches and tools evolved to support the reconstruction of software architectures. Code Crawler [13] allows reverse engineering views of the physical architecture. The tool is based on the concept of polymeric views which are bi-dimensional visualisations of metric measurement such as *Lines of Code, Number Of Methods, Complexity, Encapsulation* etc. These views help to comprehend the software system to identify eroded and problematic artefacts. The Bauhaus [14] tool offers methods to analyze and recover the software architecture views of legacy systems; it supports the identification of re-usable components and the estimation of change impact.

In reverse engineering, software clustering is often applied to produce architectural views of applications by grouping together implementation units, functions, files etc. to subsystems that relate together. Software clustering refers to the decomposition of a software system into meaningful subsystems [15]. The clustering results help to understand the system. The basic assumption driving this kind of reconstruction is that software systems are organised into subsystems characterised by high internal cohesion and loose coupling between subsystems. Therefore, most software clustering approaches reward high cohesion within the extracted modules and low coupling between the modules [16]. Barrio [17] is used for cluster dependency analysis, by using the Girvan–Newman clustering algorithm to extract the modular structure of programs. The work of Macoridis & Mitchell [18, 19, 20] identifies distinct clusters of similar artefacts based on cohesion and coupling by applying a search based cluster strategy. These approaches are appropriate if the purpose is merely the aggregation of associated artefacts into a first abstraction of the system to redraw component boundaries in software, in order to improve the level of reuse and maintainability.

Software architecture reconstruction approaches apply software clustering approaches to determine an architecture model of the system.

Chrstl & Koschke [21] depict the application of a search based cluster algorithm introduced in [22] to classify implementation units into a given conceptual architecture model. In a set of controlled experiments more than ninety percent of the implementation units are correctly classified into subsystems. The results indicate that an exclusively coupling based attraction function delivers better mapping results than the approach based on coupling and cohesion. Due to the given conceptual architecture the search space (clusters and dependencies between clusters) is distinctly reduced and a fair amount of the tiring process of assigning implementation units into subsystems is automated. However, it would be interesting if the approach is still feasibility if the erosion of the system is more pronounced. There is a high chance that with further erosion of the system that the error ratio would accumulate. Additionally, to apply the approach of Christl & Koschke [21] a conceptual architecture has to be evident to conduct the clustering. But as illustrated in the previous section and supported by various sources [7, 12, 23] current software systems face especially that the conceptual architecture is completely or at least partially lost.

The work of EAbreu et al. [16] complements the results of Christl & Koschke [21], showing that clustering based on the similarity metric and rewarding cohesion within subsystems and penalising coupling between subsystems does not provide satisfactory results which go beyond the visualisation of cohesive modules such as dependencies between modules, which would allow to model concepts as machine boundaries and encapsulation of modules.

The reconstruction of an architectural model, which can later be used as a conceptual architecture for further development is accompanied by two main problems, which cannot be solved with an approach which exclusively relies on maximises cohesion and minimising coupling based on a similarity function. The first problem is that a natural dependency flow from higher subsystems to modules of lower hierarchy levels exists. This dependency flow induces the cohesion and coupling based cluster algorithms to include artefacts of lower modules. Secondly, an architecture reconstruction is probably applied when the conceptual architecture is lost. Therefore a high degree of erosion might be evident in the physical architecture and correspondingly the assumption that a high internal cohesion and loosely coupling is evident might not be existent. Hence, to reconstruct an architectural model which fulfils the requirements of an architectural model more refined analysis techniques have to be applied. Other approaches base their analysis on non source code formations such as symbolic textual information available in comments, on class or method names, historical data (time of last modification, author) [24]. Other research includes design patterns as an architectural hint [25]. Sora et al. [8] enhance the basic cohesion and coupling based on the similarity and dissimilarity metric by introducing the concept of abstraction layers. Sora et al. [8] proposes an partitioning algorithm that orders implementation units into abstract layers determined by the direction of the dependencies. Sora et al. [8] do not include the possibility of unwanted dependencies. Therefore, architecture violating dependencies might bias the analysis and a higher degree of erosion leads to a solution with lower quality. Further evolved architecture reconstruction approaches aim to recover the layering of software systems as a more consistent documentation for development [26, 8].

Current approaches show the feasibility to reconstruct architectural views of software systems, however these approaches do not evaluate if these results are applicable to improve the understandability of the system or if the results are applicable as a conceptual architecture as part of the reflexion model. Additionally the illustrated architecture reconstruction approaches struggle to identify metrics beside cohesion and coupling to capture the quality of a conceptual architecture. The illustrated approaches do not consider that the physical architecture features a degree of erosion and deterioration which biases the reconstruction of a conceptual architecture. Current architecture reconstruction approaches create an abstract view of the physical architecture of the software system into. These abstraction views itself do not benefit a quality improvement of the system. They rather deliver a blue print for development stakeholders to understand the current implementation of the system. This enhanced understanding of the system can be utilised to conduct refactorings to improve the physical design of the system. Thus, the reconstruction of a conceptual architecture without changing the physical architecture will not improve the quality of the software system. Especially if the conceptual architecture has been reconstructed based on the source code of an eroded software system refactoring is required to realign the eroded design with the new conceptual architecture model. Hence, to improve the quality of the system the physical architecture has to be refactored in conjunction with the conceptual architecture to improve the overall design of the system.

## *2.3 Automatic Architecture Refactoring*

One of our objectives of this study is to automatically realign the physical architecture with a given or reconstructed conceptual architecture. We understand the resolution of architecture violations as a migration of the physical architecture model to a new instance of the physical architecture model which features the same behaviour but aligns to a given conceptual architecture model. Therefore, work which feature automatic refactoring, architecture realignment and migration of software systems is of particular interest.

Refactoring is the process of changing the design of a software system by preserving the behavior [27]. This design improvement should positively benefit software quality attributes [10] such as testability, modularity, extendibility, exchangeability, robustness etc. Gimnich and Winter [28] depict migration as an exclusively technical transformation with a clear target definition. The legacy system is considered as featuring the required functionality and this is not changed by applying the migration. Therefore, the refactoring of a software system can be understood as a migration of a software system to another version which fulfils other quality criteria. Hasselbring, et al. [29] describe architecture migration as the adaptation of the system architecture e.g. the migration from a monolithic system towards a multi-tier architecture. Heckel et al.[30] illustrates a model driven approach to transform legacy systems into multi-tier or SOA architecture by applying the four steps *code annotation, reverse engineering, redesign and forward engineering*. The code annotation is the manual equipment with a foreseen association of architectural elements of the target system, e.g., GUI, application logic or data conducted by the development stakeholders [30]. The remaining three stages are executed guided by the annotations. If the identified solution is not satisfying the approach is iteratively repeated.

Ivkovic & Kontogiannis [1] propose an iterative framework for software architecture refactorings as a guideline to refactor the conceptual architecture model towards *Soft Quality Goals* using model transformations and quality improvement semantic annotations. The first step of Ivkovic & Kontogiannis [1] approach requires determining a *Soft Goal* hierarchy. The Soft Goal hierarchy is a set of *Soft Goals* ordered by relevance. The *Soft Goal* model assigns metric configurations to the *Soft Goals* high maintainability, high performance and high security. In the second phase a set of candidate architectural refactorings are selected which lead to improvements towards one of the *Soft Goals*. In the third stage the derived refactorings are annotated with compatible metrics which measure quality aspects of the concerned *Soft Goal*. Metric values are determined before and after conducting the refactoring to establish if the refactoring has a positive effect onto the quality attribute of the *Soft Goal*. Finally, the refactorings are iteratively conducted by selecting each soft goal of the soft goal hierarchy and implementing the refactorings with the highest benefit based on the previous metric measurements. O'Keeffe and Cinnéide [31] propose an automatic refactoring approach to optimize a set of quality metrics. They developed a set of seven complementary pairs of refactorings to change the structure of the software system. Metaheurisitc algorithms such as *multiple ascent hill-climbing, simulated annealing* and *genetic algorithm* are then used to apply the implemented refactorings. The fitness function to evaluate the refactored source code instance employs an implementation of the Bansiya's QMOOD hierarchical design quality model [32]. The QMOOD model comprises eleven weighted metrics depending on the weighting of these metrics the software quality attribute understandibility, reusability and flexibility can be expressed as a numerical measurement [32]. O'Keeffe and Cinnéide [31] utilizes these three different weightings as different fitness functions to refactor a system towards the desired quality attributes. They found that some of the example projects can be automatically refactored to improve quality as measured by the QMOOD evaluation functions. The variation of weights on the evaluation function has a significant effect on the refactoring process. The results show that first-ascent hill climbing produces significant quality improvements for the least computational expenditure, steepest-ascent hill climbing delivered the most consistent improvements and the simulated annealing implementation is able to produce the greatest quality improvements with some examples. O'Keeffe and Cinnéide [31] go on to state that the output code of the flexibility and understandability produced meaningful outputs in favour of the desired quality attributes where the reusability function was not found to be suitable to the requirements of search-based software maintenance because the optimal solution includes a large number of featureless classes.

## 3. An Architecture Reconstruction and Refactoring Framework

This section describes an automatic architecture reconstruction and transformation process designed to support the reconstruction of a conceptual architecture model of a software system and the migration of the analysed software system towards a given conceptual architecture model.

In the previous section a variety of architecture reconstruction, refactoring and migration approaches have been reviewed. It has been shown that current architecture reconstruction approaches are feasible to extract views of the physical architecture. The reconstructed architectural views can help development stakeholders to understand the current design of the system. However, the approaches are not aiming to reconstruct a conceptual architecture of the system or a blue print of the system which can be used for further development and compliance checking. Consequently, the re-creation of a conceptual architecture remains a tedious manual process which requires analyzing domain and technical environment aspects in compliance with the evident software system. One of the main problems while reconstructing a conceptual architecture is the erosion which might be evident in the system and bias the extraction of a conceptual architecture. The identification of violating dependencies is hard due to the uncertainty of the system deterioration. Automatic refactoring approaches refactor architectural views [1] or the source code [31] of the system towards predefined quality goals. These quality goals are represented as combinations of metrics which measure the desired quality goal. Migration and transformation approaches transform legacy systems into multi-tier or SoA architectures [29]. Most approaches require a substantial part of manual classification [30] hence a good understanding of the system is required. Other approaches transform views of the architecture without transforming the source code of the system [28]. Furthermore, approaches either transform or reconstruct architectural views or change the source code of the system. To our current understanding none of the reviewed approaches aim to provide a conceptual architecture view as well as a corresponding physical architecture model. We believe that a conceptual architecture model as well as a violation free physical model is one of the key requirements to enable an active and continuous architecture management. Additionally, the evidence of a corresponding reflexion model delivers the base for further development and refactoring of the system towards better architectural design.

Based on this we propose and evaluate a combination of architecture reconstruction techniques to extract a conceptual architecture model and refactoring techniques to obtain an aligned physical architecture and conceptual architecture model. The process reflects that a conceptual architecture model can be based on acknowledged software design principles represented as architecture styles, design patterns and software metrics. The conceptual architecture model represents a target definition for the migration of the physical architecture model. The reconstruction and transformation process is outlined in Figure 1 which illustrates the input and output relationships.

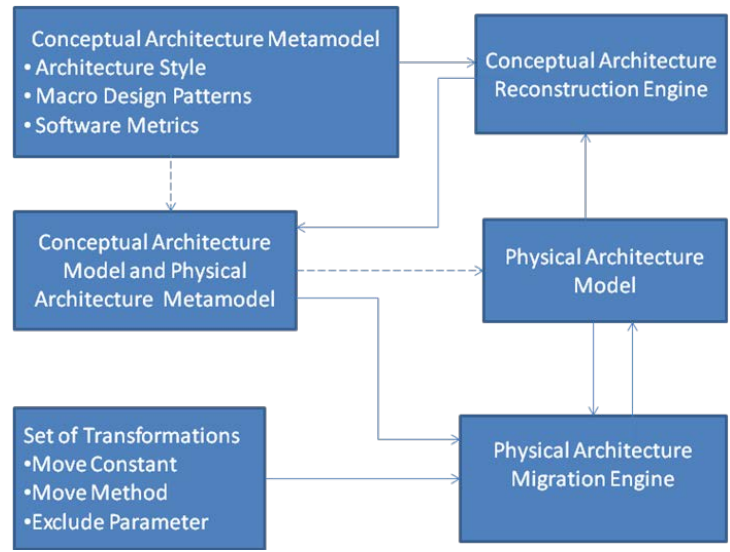

**Fig.** 1 Automatic Architecture and Migration Framework

A prototype has been developed to enable an evaluation of the feasibility of the framework. This prototype allows the reconstruction of a conceptual architecture as well as the refactoring of the physical architecture towards this conceptual architecture model on the basis of java software systems. Dependency analysis as well as the migration of source code instances are enabled by applying the RECODER source code optimisation and transformation framework [33]. The following two sections illustrate architecture reconstruction and architecture migration as implemented in our framework.

## 3.1 Architecture Reconstruction

An automatic conceptual architecture reconstruction framework is useful if the desired architecture of a system is lost. As previously stated, the manual reconstruction of conceptual architecture in fairly eroded systems is complex and labour intensive [34]. The objective of this component is to evaluate if the automatic reconstruction of a reflection model is feasible in terms of delivering a modular structure of a software system as a basis for further development. The reconstructed architecture model delivers a conceptual blueprint of the system that implements established design principles of software architectures. This blueprint can be used in further development, refactoring and architecture monitoring.

Considering the necessity to apply an architecture reconstruction it can be assumed that no or only limited documentation of the conceptual architecture exists and the physical architecture of the system is eroded. Hence, regular compliance checking has not been conducted due to the missing basis for comparison. Additionally, it is also not possible to determine the degree of erosion as the basis for comparison in the form of a defined design description is missing and the erosion is influenced by a variety of factors such as development activity, design drift, design understanding of development stakeholders, framework changes and time without compliance checking. Additionally, the requirements of an ideal conceptual architecture candidate change over time caused by requirement, environment and framework changes. The definition of an ideal conceptual architecture depends on variables such as the development environment, development philosophy, applied frameworks and functional requirements. It is hard to capture all these variables within an automated approach based on the analyses of legacy code. However, we are convinced that at least having a conceptual architecture has long term benefits on the life cycle and quality of the software system.

We suggest a search based cluster approach to reconstruct a conceptual architecture. This decision is based on the complexity of the problem, the size of the search space and also the multiplicity of optimal solutions. To date the reconstruction of a layered architecture style with transparent horizontal layers has been implemented. A search based clustering algorithm similar to the clustering approach of Mitchell and Mancoridis [20] classifies the implementation units into $n$ layers. As an acknowledged software clustering concept the clustering penalizes high coupling between clusters and rewards high cohesion within the clusters.

We employ a greedy metaheurisitic to identify a start solution and apply a steepest ascent hill climbing metaheuristic to improve this initial solution. Our approach utilizes the work of Harman [35] which states that metrics can act as fitness functions. Our objective is to recreate a system architecture that exhibits good modularity.

We designed a fitness function *Solution Quality* to evaluate the fitness of a solution. Based on the *Soft Goal* graph of Ivkovic & Kontogiannis [1] we utilize the *Coupling Between Objects* metric as measurement for modularity. Every dependency between implementation units is annotated with a *CBO* measurement. Our greedy algorithm classifies the implementation units into clusters based on rewarding cohesion and penalizing coupling. The clusters are ordered ascending based on the ratio of incoming and outgoing dependencies. Additionally, we reward solutions with more layers. The *Solution Quality* is multiplied with the number of layers in the system. However, at this stage we only allow solutions with three or less layers. The steepest ascent hill climbing algorithm tries to increase the *Solution Quality* measurement by swapping implementation units between clusters.

In our model an architecture violation is a dependency from a lower layer to a higher layer. These dependencies deteriorate the encapsulation of two layers if we take the conceptual architecture as an optimal solution. As the system probably features an indefinite degree of deterioration we do not just want to minimize the number of architecture violations. Hence, just relying on the minimization of architecture violations would model the deteriorated system into a conceptual model and therefore not challenge an improvement of the system design. Our overall aim is to obtain a violation free architecture of the system. To support this approach we classify between violations which can be resolved by the automatic refactoring (defined in sections 3.2.1 and 3.2.2) and violations our approach is not capable to resolve. Each dependency is tested if it can be resolved by one of our three automatic refactoring transformations. If a dependency can be resolved by the application of refactoring the CBO weight of the dependency is multiplied with a factor of 0.25 and therefore the dependency is rather an accepted architecture violation as it does only increase the coupling between layers by a small degree. Hence, we penalize the inclusion of architecture violations which cannot be resolved with our by multiplying the CBO measurement with a factor of 2.0, which strongly increases the coupling between layers and therefore penalizes the solution.

The output of this conceptual architecture reconstruction is a conceptual architecture model which comprises ordered layers and implementation units which are mapped into these layers. Therefore, a reflexion model has been created. However, the physical architecture might feature dependencies which violate with the reconstructed conceptual architecture model [9].

## 3.2 Automatic Architecture Refactoring

This section describes an automatic refactoring framework to migrate the physical architecture towards a given conceptual architecture by applying a set of transformations.

Our automatic refactoring approach expects as input the reflexion model of a software system. This reflexion model can be user-defined or can be created by our previously illustrated architecture reconstruction method. The objective of this component is to deliver an automatic source code transformation approach that has the potential to re-establish the modularity of eroded software.

The migration framework aims to resolve architecture violations which cannot be resolved by reclassifying implementation units into different subsystems of the conceptual architecture model. The origin of these architecture violations is hidden in the implementation of the system, which does not align with the conceptual modularity and decomposition of the system [7]. To resolve architecture violations of this kind the source code of the implementation unit has to be migrated to comply with the conceptual architecture. A set of allowed transformations had to be specified to migrate the system from one instance to another.

The objective of the automatic refactoring is the resolution of unwanted dependencies between implementation units. This requires the application of refactorings which alter the dependency structure between implementation units. Rosik, Le Gear, Buckley, Babar and Connolly [2] found that a significant number of violations are based on misplaced functionality. Another common reason for the erosion of design is the injection of higher-classified implementation units as a parameter and access to the structure and behaviour of these objects from lower-classified implementation units [12]. Therefore the three transformations

*move method*, *move constant* and *exclude parameter* have been implemented within our proof-of-concept prototype. These transformations refactor the implementation unit which causes the architecture violation as well as the interrelated callers of the refactored code element. The behaviour of these transformations is as follows:

### 3.2.1 Move Method and Move Constant

These transformations move a method or constant that causes an architecture violation to any other implementation unit. As it cannot be assumed to which implementation unit the refactored code artefact should be moved, the code artefact in question is placed into every implementation unit of the system. For each of these outcomes a new instance of the source code is created as the base for the application of further transformations.

### 3.2.2 Exclude Parameter

The exclude parameter transformation excludes one parameter of a method. The code elements which reference or access the parameter are moved to the caller implementation units. Currently the order of execution can be changed by applying this transformation and consequently the program behaviour might change. Our current solution is to exclude the parameter and instead include a listener pattern which notifies the top layer implementation unit to execute the refactored code elements. Based on an identified architecture violation one of the three implemented transformations can be selected. A new instance of the software system is created based on the application of every transformation. The complexity of an exhaustive search would quickly result in an uncontrollable number of generated software system instances. Figure 2 illustrates the uncontrolled generation of software system instances.

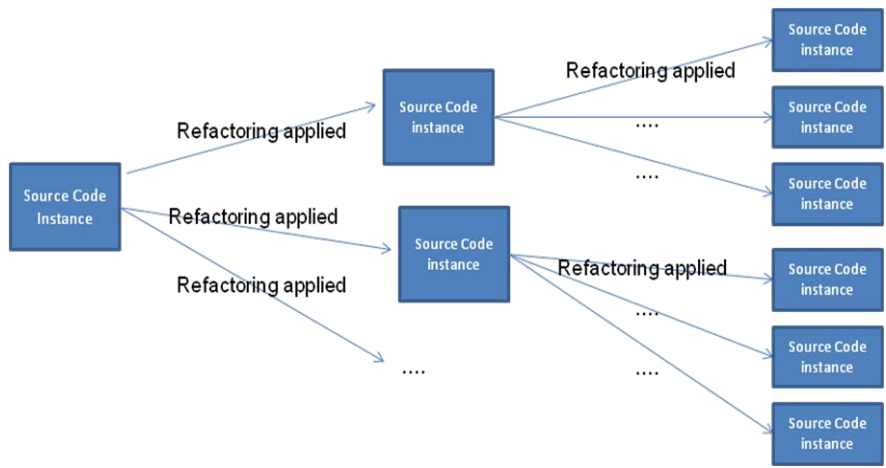

Figure 2: Evolution of Source code instances based on applied transformations

Due to this computational complexity we apply a greedy algorithm to control the reproduction process of software system instances. Based on the initial solution a population of new software instances is created. The fittest solution is selected based on the lowest number of architecture violations as the primary selection criteria. If two solutions feature the same number of architecture violations the selection is based on the fitness function illustrated in section 3.1. Figure 3 illustrates the selection strategy and reproduction for two generations.

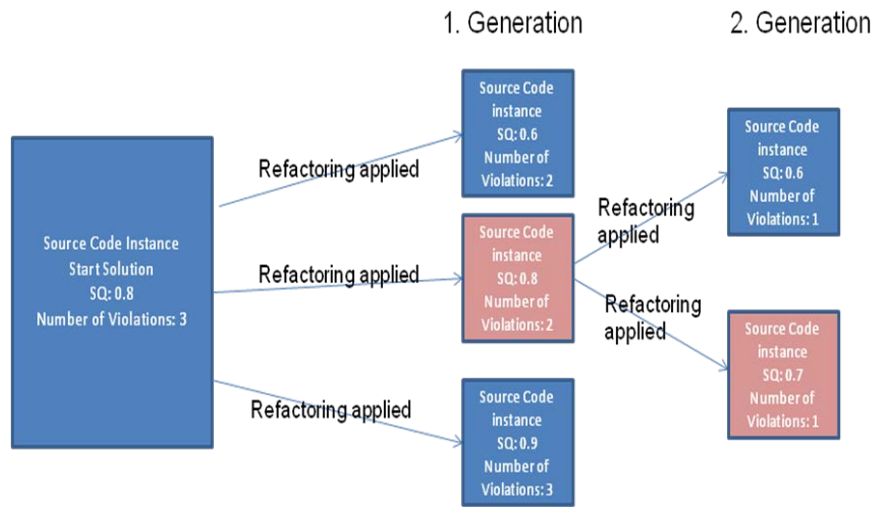

**Figure 3 : Selection Strategy of Greedy algorithm**

## 4. Evidence of Feasibility and Evaluation

Our initial evaluation of the prototype has utilised controlled experiments. Evaluation of the framework is based on the architecture reconstruction of a small self-developed system (comprising 15 implementation units) which follows a MVC architecture style. The system is structured such that in each case five of the implementation units fulfil model, view or controller functionality. If we measure the fitness for a self defined optimal conceptual MVC- architecture model with no erosion in the physical design we measure a *Solution Quality* of 0.66185.

### 4.1 Reconstruction of Conceptual Architecture

A set of 11 experiments has been conducted, in each experiment an architecture violating dependency is added and then resolution is attempted. These imposed architecture violations conflict with the MVC architecture. The type of violation is equally distributed by adding wrongly placed methods, constants and wrongly injected parameters. The results of the experiments are shown in Ta**ble 1.**

Table 1 Results Architecture Reconstruction Experiment

| No. Injected Architecture Violations | No. Layers | Misplaced implementation units | Architecture violations | Solution Quality |
|---|---|---|---|---|
| 0 | 3 | 2 | 0 | 0.64835 |
| 1 | 3 | 2 | 0 | 0.63464 |
| 2 | 3 | 2 | 1 | 0.53571 |
| 3 | 3 | 3 | 2 | 0.55833 |
| 4 | 3 | 3 | 3 | 0.45889 |
| 5 | 3 | 3 | 3 | 0.42652 |
| 6 | 2 | 5 | 2 | 0.58534 |
| 7 | 2 | 5 | 2 | 0.58534 |
| 8 | 2 | 6 | 3 | 0.54532 |
| 9 | 2 | 6 | 5 | 0.53345 |
| 10 | 2 | 6 | 6 | 0.52345 |

The results show that the prototype is able to identify a conceptual architecture model of the system and classify the implementation units into corresponding layers. The *Solution Quality* as a fitness representation tends towards lower values with increasing erosion of the system. The only break in this general trend is the reduction of layers in the conceptual architecture to two which causes disjointed values.

In each of the experiments the resulting conceptual architecture features a set of implementation units which are not classified correctly and also the number of identified violations differs from the number of initiated architecture violations; hence the identified architecture violations are not necessarily identical to the introduced architecture violations. The experiments also show that, based on the rising number of misplaced implementation units, the constructed conceptual architecture model drifts further from the initial MVC architecture. However, at this stage the suggestion of a conceptual architecture model which can be used as a base for further development and refactoring to regain a degree of system modularity seems to be feasible.

## *4.2 Realignment of Physical Architecture with Conceptual Architecture Model*

We conducted a second set of experiments based on the physical architecture of our self developed MVC example. We utilised the physical design with 10 injected architecture violations and the reconstructed conceptual architecture model with 2 layers from our previous experiments. The objective of these experiments is to evaluate the automatic refactoring of the physical architecture towards a given conceptual architecture model.

To evaluate the feasibility of the automatic refactoring towards a given conceptual architecture it is necessary to evaluate if the process contributes to a quality improvement of the software system. The proposed approach aims to re-establish the modularity of the system. The main objective of the automatic refactoring is the reduction of architecture violations.

However, the number of architecture violations depends strongly on the given conceptual architecture model. So far the *Solution Quality* fitness function is available to evaluate the modularity of the system. Table 3 shows the results of this experi**ment.**

Table 2 Results Architecture Refactoring Experiment

| Generation | Number of Architecture Violations | Solution Quality |
|---|---|---|
| 1 | 6 | 0.52345 |
| 2 | 5 | 0.50432 |
| 3 | 4 | 0.54545 |
| 4 | 3 | 0.56362 |
| 5 | 2 | 0.53756 |
| 6 | 2 | 0.53756 |

A reduction of architecture violations can be observed during the first five generations. From the fifth generation no appropriate move can be identified to resolve the remaining architecture violation. The *Solution Quality* fitness function measurement reflects no significant quality improvement of the refactored system and no clear trend of the *Solution Quality* measurement can be recognised. The reason for this might be the individual evaluation of architecture violations in the model in respect to their resolvability with our implemented refactorings. To evaluate the quality of the generated solution more general metrics should be applied to allow estimating the overall quality development of the system.

In general, it has been found that violations based on wrongly placed constants can be completely resolved. The outcome of resolutions using the move method and exclude parameter transformations depends on the dependencies of the method and parameter to the initial containing implementation unit. If no interrelation to the containing implementation unit exists the method can be placed into other implementation units or the parameter can be excluded and the violation resolved. However, these preliminary results show that a migration from one instance of a software system to another is feasible by applying a set of defined transformations which align the software system with a given conceptual architecture model.

## 5 Limitations

The conducted evaluation is preliminary but is encouraging. Further application in real scenarios is necessary (and is ongoing) to more fully assess the applicability of our Architecture Reconstruction and Migration Framework.

## 6 Conclusions and Future Work

This paper describes a framework designed to reconstruct a conceptual architecture model for legacy systems and to migrate the physical architecture model of legacy systems towards a given conceptual architecture model. Based on the theoretical illustration of the causes and consequences of deteriorated software design, the possibility to utilise the conceptual architecture model as a metamodel for the physical architecture is illustrated. The method of operation of the architecture reconstruction by utilizing acknowledged macro-architecture design principles and the physical architecture model is described. Furthermore the principles of operation of the software migration framework by utilising the conceptual architecture model, applying design patterns, software metrics and source code transformation are described. Finally, preliminary results of our feasibility evaluation are presented and discussed.

At this time our prototype addresses a limited set of architecture styles and transformations. We are working to extend the number of possible architecture styles by introducing vertical layering and impervious layers to model functional decomposition and machine boundaries in the conceptual architecture. Further research will also focus to resolve architecture violations by the migration of the source model towards design patterns. Another current working area is the extension of the *move method* and *exclude parameter* transformations to migrate interrelations to containing implementation units. The current search strategies are immature. It will be beneficial especially for the evaluation of larger software systems to guide the search towards more promising solution candidates by applying other search strategies e.g *genetic algorithms*. Future work will involve evaluating if the migration approach has the potential to migrate a software system from one conceptual architecture model to another. This is of particular interest if the conceptual architecture of a system changes due to requirement, environment and technology changes and the conceptual architecture model and mapping of implementation units into this new architecture can be defined.